\documentstyle[sprocl,epsf]{article}

\def\Journal#1#2#3#4{{#1} {\bf #2}, #3 (#4)}

\def\NP{{\em Nucl.~Phys.}}
\def\PL{{\em Phys.~Lett.}}
\def\PR{{\em Phys.~Rev.}}
\def\PRL{\em Phys.~Rev.~Lett.}
\def\ZP{\em Z.~Phys.}
\def\PTP{\em Prog.~Theor.~Phys.}

\newcommand{\bra}[1]{\langle #1 |}
\newcommand{\ket}[1]{|#1\rangle}

\newcommand{\ab}{{\alpha\beta}}
\newcommand{\tr}{{\rm tr}}
\newcommand{\T} {\mbox{T}}

\def\la{\Lambda_{\rm QCD}}

\begin{document}

\title{\hfill {\footnotesize RIKEN-BNL PREPRINT} \\ 
\vskip 5mm
LONG--RANGE INTERACTIONS \\ OF SMALL COLOR DIPOLES\footnote{
Invited talk at the Third ``Continuous Advances in QCD" Workshop dedicated 
to the memory of V.N. Gribov; Minneapolis, April 16--19, 1998.}}

\author{H.~Fujii and D.~Kharzeev}

\address{RIKEN-BNL Research Center,\\
Brookhaven National Laboratory,\\
Upton, NY 11973, USA}

\maketitle
\abstracts{ 
We study the scattering of small color dipoles (e.g., heavy quarkonium
states) at low energies. We find that even though the couplings of
color dipoles to the gluon field can be described in perturbation
theory, at large distances the interaction becomes totally
non--perturbative. The structure of the scattering amplitude, however,
is fixed by the (broken) chiral and scale symmetries of QCD; the
leading long--distance contribution arises from the correlated
two--pion exchange. We use the spectral representation technique to
evaluate both perturbative and non-perturbative contributions to the
scattering amplitude. Our main result is the sum rule which relates
the overall strength of the non--perturbative interaction between
color dipoles to the energy density of QCD vacuum.}

\section{Introduction}

In a 1972 article entitled ``Zero pion mass limit in interaction at
very high energies" \cite{AG}, A.A. Anselm and V.N. Gribov posed an
interesting question: what is the total cross section of hadron
scattering in the chiral limit of $m_{\pi} \to 0$? On one hand, as
everyone believes since the pioneering work of H. Yukawa, the range of
strong interactions is determined by the mass of the lightest meson,
i.e. is proportional to  $\sim m_{\pi}^{-1}$. The total cross sections
may then be expected to scale as $\sim m_{\pi}^{-2}$, and would tend
to infinity as $m_{\pi} \to 0$. On the other hand, soft--pion
theorems, which proved to be very useful in understanding low--energy
hadronic phenomena, state that hadronic amplitudes do not possess
singularities in the limit $m_{\pi} \to 0$, and one expects that the
theory must remain   self-consistent in the limit of the vanishing
pion mass.

At first glance, the advent of QCD has not made this problem any
easier; on the contrary, the presence of massless gluons in the theory
apparently introduces another long--range interaction. In this
article, we try to address this problem considering the scattering of
small color dipoles. The reason for choosing this, somewhat specific,
example is simple: asymptotic freedom dictates that strong coupling
becomes weak at short distances, and since the small size of dipoles
$r$ introduces a natural infrared cut-off $r \ll \la^{-1}$, one
expects that their interactions can be systematically treated in
QCD. This approach was pursued both at low
\cite{Voloshin,Peskin,AK,LMS,KS} and high \cite{Mueller} energies.

It has been found, however, that at sufficiently high energies the
perturbative description of ``onium--onium" scattering breaks down
\cite{Mueller1,Dok}. The physical reason for this is easy to
understand: the higher the energy, the larger impact parameters
contribute to the scattering, and at large transverse distances the
perturbation theory inevitably fails, since the virtualities of
partons in the ladder diffuse to small values (``Gribov
diffusion"). At this point, the following questions arise: Does this
mean that the problem becomes untreatable? Does the same difficulty
appear at large distances in low--energy scattering?  And, finally,
what (if any) is the role played by pions? 

\section{Perturbation Theory}
\label{sec:pert}

The Wilson operator product expansion \cite{Wilson} allows one to
write down the scattering amplitude (in the Born approximation) of two
small color dipoles in the following form\cite{Peskin}:
\begin{eqnarray}
V(R) &=& -i \int dt  
\bra{0} \T \left( \sum_i c_i O_i (0)\right) 
           \left( \sum_j c_j O_j (x)\right) 
\ket{0}, 
\label{ope1}
\end{eqnarray}
where $x = (t, R)$, $O_i(x)$ is the set of local gauge-invariant
operators expressible in terms of gluon fields, and $c_i$ are the
coefficients which reflect the structure of the color dipole. At small
(compared to the binding energy of the dipole) energies, the leading
operator in (\ref{ope1}) is the square of the chromo-electric field
$(1/2) g^2 {\bf E}^2$ \cite{Voloshin,Peskin} --- other twist--two
operators contain covariant derivatives leading to the powers of
momentum in the amplitude and are therefore suppressed at small
energies. 

Keeping only this leading operator, we can rewrite (\ref{ope1}) in
a simple form   
\begin{eqnarray}
V(R) &=& 
- i\Big ( \bar d_2 \frac{a_0^2}{\epsilon_0} \Big)^2 
\int d t  
\bra{0} \T \ \frac{1}{2}g^2 {\bf E}^2 (0)\ 
\frac{1}{2} g^2 {\bf E}^2(t,R) \ket{0}, 
\label{pot1}
\end{eqnarray}
where $\bar d_2$ is the corresponding Wilson coefficient defined by
\begin{equation}
\bar d_2 \frac{a_0^2}{\epsilon_0}
=\frac{1}{3N}\bra{\phi} r^i \frac{1}{H_a + \epsilon} r^i \ket{\phi},
\end{equation}
where we have explicitly factored out the dependence on the quarkonium
Bohr radius $a_0$ and the Rydberg energy $\epsilon_0$; $N$ is the
number of colors, and $\ket{\phi}$ is the quarkonium wave function,
which is Coulomb in the heavy quark limit\footnote{The Wilson
coefficients $\bar d_2$, evaluated in the large $N$ limit, are
available for $S$ \cite{Peskin} and $P$ \cite{DK} quarkonium states.}.
The factors $a_0$ and $\tau\sim 1/\epsilon_0$ represent the
characteristic dimension and fluctuation time of the color dipole,
respectively. The approximate expression (\ref{pot1}) is justified
when the wavelength of gluon fields is large compared to $a_0$ and
they change slowly compared to $1/\epsilon_0$. 

In physical terms, the structure of (\ref{pot1}) is transparent: 
it describes the elastic scattering of two dipoles which act on 
each other by chromo-electric dipole fields; color neutrality permits 
only the square of dipole interaction in the elastic scattering.

The amplitude (\ref{pot1}) was evaluated before \cite{Peskin} 
in perturbative QCD using functional methods. For our purposes,
however, it is convenient to use a different technique based on the
spectral representation. As a first application of this approach, we
will reproduce the result of \cite{Peskin} by a different method. 

It is convenient, as a first step, to express $g^2 {\bf E}^2$ in terms
of the gluon field strength tensor \cite{NS}:
\begin{equation}
g^2 {\bf E}^2
=
- \frac{1}{4}g^2    G_{\alpha\beta}G^{\alpha\beta}
+g^2(- G_{0\alpha} G_0^\alpha
+\frac{1}{4} g_{00} G_{\alpha\beta}G^{\alpha\beta})
=
 \frac{8 \pi^2}{b} \theta_\mu^\mu + g^2 \theta_{00}^{(G)}  
 \label{e2}
\end{equation}
where
\begin{eqnarray}
\theta_\mu^\mu \equiv
\frac{\beta(g)}{2g} G^{\alpha\beta a} G_{\alpha\beta}^{a} =
- \frac{b g^2}{32 \pi^2} G^{\alpha\beta a} G_{\alpha\beta}^{a} 
 \ . \label{trace}
\end{eqnarray}
Note that as a consequence of scale anomaly \cite{scale}, 
$\theta_\mu^\mu$ is the trace of the energy-momentum tensor 
of QCD in the chiral limit of vanishing light quark masses, 
and the $\beta$ function in (\ref{trace}) does not contain the
contribution of heavy quarks \cite{dec} ({\it i.e.}
$b=\frac{1}{3}(11N-2N_f)=9$).

Let us now write down the spectral representation for the correlator 
of the trace of energy-momentum tensor:
\begin{eqnarray}
\bra{0} \T \theta_\mu^\mu(0) \theta_\mu^\mu(x) \ket{0} = 
\int d \sigma^2 \rho_\theta (\sigma^2) \Delta_F(x;\sigma^2),
\label{spectral}
\end{eqnarray}
where the spectral density is defined by
\begin{equation}
\rho_\theta (k^2) \theta(k_0) =
\sum_n (2\pi)^3 \delta^4(p_n-k)|\bra{0}\theta_\mu^\mu\ket{n} |^2 ,
\label{def_spect}
\end{equation}
and
\begin{eqnarray}
i\Delta_F(x;\sigma^2) 
&=&
i \int \frac{d^4k}{(2\pi)^3}\delta(k^2-\sigma^2)\theta(k_0)
(e^{-ikx} \theta(x^0)+e^{ikx} \theta(-x^0))
\end{eqnarray}
is the Feynman propagator of a scalar field.

Using the representation (\ref{spectral}) in (\ref{pot1}), we get
\begin{eqnarray}
V_\theta(R) &=& 
-i \Big ( \bar d_2 \frac {a_0^2}{\epsilon_0} \Big )^2 
\Big ( \frac{4 \pi^2 }{b}\Big ) ^2
\int dt 
\int d \sigma^2  \rho_\theta(\sigma^2) \Delta_F(x;\sigma^2)
\nonumber \\
&=& -
\Big ( \bar d_2 \frac {a_0^2}{\epsilon_0} \Big )^2 
\Big ( \frac{4 \pi^2 }{b}\Big ) ^2
\int d \sigma^2  \rho_\theta (\sigma^2)
\frac{1}{4\pi R}e^{-\sigma R}.
\label{yukawa}
\end{eqnarray}
We see that the potential (\ref{yukawa}) can be represented as a 
superposition of Yukawa potentials corresponding to the exchange of
scalar quanta of mass $\sigma$. 

Our analysis so far has been completely general; the dynamics 
enters through the spectral density (\ref{def_spect}). Let us first
evaluate this quantity in perturbation theory. In this case we have to
evaluate the contribution of two--gluon states (see Fig.~1(a)) defined
by
\begin{eqnarray}
\rho_\theta^{\rm pt}(q^2) & \equiv  &
\sum (2\pi)^3 \delta^4 (p_1+p_2-q) 
| \bra{0} \theta^\alpha_\alpha 
\ket{p_1 \varepsilon_1 a, p_2\varepsilon_2 b}|^2 , \label{pertd}
\end{eqnarray}
where the phase-space integral should be understood as well as the
summations over the polarization and the color indices of the gluons.

The evaluation of (\ref{pertd}) is straightforward: for $SU(N)$, one
has
\begin{eqnarray}
\rho^{\rm pt}_{\theta}(q^2) &=& 
\left ( \frac{bg^2}{32\pi^2} \right )^2\frac{N^2-1}{4\pi^2} q^4  .
\label{pdens}
\end{eqnarray}
Then, substituting (\ref{pdens}) into (\ref{yukawa}) and performing 
the integration over invariant mass $\sigma^2$, we get, for $N=3$
\begin{eqnarray}
V_\theta(R) &=&
-  g^4 \Big ( \bar d_2 \frac{a_0^2}{\epsilon_0}\Big)^2
\frac{15}{8\pi^3}\frac{1}{R^7}.
\label{casimir}
\end{eqnarray} 

The $\propto R^{-7}$ dependence of the potential (\ref{casimir}) is a
classical result known from atomic physics \cite{cas}; as is apparent
in our derivation (note the time integration in (\ref{yukawa})), the
extra $R^{-1}$ as compared to the Van der Waals potential $\propto
R^{-6}$ is the consequence of the fact that the dipoles we consider
fluctuate in time, and the characteristic fluctuation time $\tau \sim
\epsilon_0^{-1}$, is small compared to the spatial separation of the
``onia'' : $\tau \ll R$. 

Finally we note that in perturbation theory 
$\theta_\mu^\mu$ is of order $g^2$,
(even though the matrix element of the $\theta_\mu^\mu$, as will be 
discussed below, is in general non--perturbative), and accordingly the
potential (\ref{casimir}) has the prefactor $g^4$. Then the second
term in (\ref{e2}) gives the contribution of the same order in $g$;
this contribution is due to the tensor $2^{++}$ state of two gluons
and can be evaluated in a completely analogous way. Adding this
contribution to (\ref{casimir}), we reproduce the result of
Ref.~\cite{Peskin}: 
\begin{eqnarray}
V(R) &=&
-  g^4 \Big ( \bar d_2 \frac{a_0^2}{\epsilon_0}\Big)^2 
\frac{23}{8\pi^3}\frac{1}{R^7}; \label{pertres}
\end{eqnarray} 
note that our $\bar d_2$ is related to the $d_2$ of Ref.~\cite{Peskin} by
$ d_2 a_0 \epsilon_0 = \bar d_2 g^2 $.

\begin{figure}[tb]
\epsfxsize=0.5\textwidth
\centerline{\epsffile{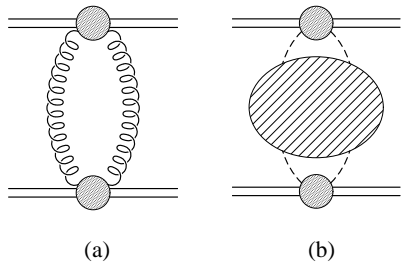}}
\caption{Contributions to the scattering amplitude from (a) two
gluon exchange and (b) correlated two pion exchange.}     
\end{figure}

\section{Beyond the Perturbation Theory: The Role of Pions}
\label{sec:beyond}

At large distances, the perturbative description breaks down, because, 
as can be clearly seen from (\ref{yukawa}), the potential becomes
determined by the spectral density at small $q^2$, where the
transverse momenta of the gluons become small. To see this explicitly
in the dispersive language, let us consider the correlator
\begin{equation}
\Pi(q^2) = \int d^4 x e^{iqx}
\bra{0} \T\theta_\mu^\mu(x)\theta_\mu^\mu(0)\ket{0}
=
\int d\sigma^2 \frac{\rho_\theta (\sigma^2)}{\sigma^2-q^2-i\epsilon}.
\end{equation}
An important low--energy theorem \cite{nsvz} states that, as a
consequence of the broken scale invariance, 
\begin{equation}
\Pi(0)=-4\bra{0} \frac{\beta(g)}{2g}G^{\ab a}G_\ab^a(0)\ket{0}. \label{condens}
\end{equation}
The operator on the r.h.s.\ of (\ref{condens}) is regularized by
subtracting the contribution of perturbation theory. The vacuum
expectation value of this operator therefore measures the energy
density of non--perturbative fluctuations in the QCD vacuum
\cite{SVZ}. The low--energy theorem (\ref{condens}) therefore implies
the sum rule for the spectral density \cite{nsvz}:  
\begin{equation}
\int \frac{d\sigma^2}{\sigma^2}
[\rho_\theta^{\rm phys}(\sigma^2)-\rho_\theta ^{\rm pt}(\sigma^2)]
=-4\bra{0}\frac{\beta(g)}{2g}G^{\ab a}G_\ab^a\ket{0}
=-16\epsilon_{\rm vac}
\ne 0 ,
\label{let}
\end{equation}
where $\rho_\theta ^{\rm pt}(\sigma^2)$ is given by (\ref{pdens}), 
and the vacuum energy density is 
$\epsilon_{\rm vac} = (1/4) \langle \theta_\mu^\mu \rangle \simeq - 
(0.24 \mbox{ GeV})^4$ \ \cite{SVZ}. 
Since the physical spectral density, $\rho_\theta^{\rm phys}$, should
approach the perturbative one at high $\sigma^2$, the integral in
(\ref{let}) can accumulate its value required by the r.h.s. only in
the region of relatively small $\sigma^2$\ \footnote{The analysis of
sum rules shows however that the approach to the asymptotic freedom in
the scalar channel is rather slow \cite{SVZ}.}.  

At small invariant masses, we have to saturate the physical spectral
density in the sum rule (\ref{let}) by the lightest state allowed in
the scalar channel -- two pions:
\begin{eqnarray}
\rho_\theta^{\rm phys} (q^2) 
& =  &
\sum (2\pi)^3 \delta^4 (p_1+p_2-q) 
| \bra{0} \theta^\alpha_\alpha 
\ket{\pi(p_1) \pi(p_2)}|^2 , \label{physpi}
\end{eqnarray} 
where, as in (\ref{pertd}), the phase--space integral is understood. 

Since, according to (\ref{trace}), $\theta^\alpha_\alpha$ is gluonic
operator, the evaluation of (\ref{physpi}) requires the knowledge of
the coupling of gluons to pions. This is a purely non--perturbative
problem, but it can nevertheless be rigorously solved, as it was shown
in Ref.~\cite{VZ} (see also \cite{NS}). The idea of \cite{VZ} is the
following: at small pion momenta, the energy--momentum tensor can be
accurately computed using the low--energy chiral Lagrangian: 
\begin{equation}
\theta_\mu^\mu =
-\partial_\mu \pi^a \partial^\mu\pi^a +2 m_\pi^2 \pi^a \pi^a + \cdots 
\label{trl}
\end{equation}
Substituting this expression into (\ref{physpi}), in the chiral limit 
of vanishing pion mass one gets an elegant result \cite{VZ} 
\begin{equation}
\bra{0} \frac{\beta(g)}{2g} G^{\alpha\beta a} G_{\alpha\beta}^{a} 
\ket {\pi^+\pi^-} = q^2. \label{vz}
\end{equation}

The result (\ref{vz}) can actually be generalized for the coupling of
gluons to {\it any} number of pions. Indeed, consider the Lagrangian
of non--linear $\sigma$ model, 
\begin{equation}
{\mathcal{L}} = {\frac{f_{\pi}^2}{4}}\ 
\tr\ \partial_{\mu} U \partial^{\mu}U^{\dagger}
 \ - \ \Lambda\ \tr \left(M U + U^{\dagger} M^{\dagger} \right),
\label{nlsig}
\end{equation}
where $U = \exp\left( 2i \pi / f_{\pi} \right)$,  
$\pi \equiv \pi^a T^a$, $\tr\ T^a T^b = \frac{1}{2} \delta^{ab}$, $M$
is the quark mass matrix, and $\Lambda = m_\pi^2 f_{\pi}^2 / \hat{m}$,
with $\hat{m}$ being the average light quark mass. Using the
mathematical identity for a generic matrix $A$, 
\begin{equation}
\partial^{\mu} [\exp(A)] = \int_0^1 ds\ \exp(s A)\ 
\partial^{\mu} A\ \exp((1-s) A),
\end{equation}
one can explicitly evaluate the trace of the energy--momentum tensor for 
the Lagrangian (\ref{nlsig}), with the result
\begin{equation}
\theta_\mu^\mu = - 2\ \frac{f_{\pi}^2}{4}\ \tr\ \partial_{\mu} U \partial^{\mu} 
U^{\dagger} \ + \ 4 \Lambda\ \tr \left(M U + U^{\dagger} M^{\dagger} 
\right). \label{trnl}
\end{equation}
Expanding (\ref{trnl}) in powers of pion field, one can generalize
(\ref{trl})  for any (even) number of pions; to lowest order, we
reproduce (\ref{trl}). 

Now that we know the coupling of gluons to the two pion state, the
pion--pair contribution to the spectral density (\ref{physpi}) can be
easily computed by performing the simple phase space integration, with
the result   
\begin{equation}
\rho^{\pi\pi}_\theta(q^2) = \frac{3}{32\pi^2} q^4 . \label{ppi}
\end{equation}  
Multi--pion contribution can be evaluated using (\ref{trnl}); its
influence will be discussed elsewhere. However the dominant
contribution at small invariant masses $\sigma$, in which we are
primarily interested here, comes from the $\pi \pi$ state. 

Coming back to the initial expressions (\ref{pot1}), (\ref{e2}) we
find that to complete our derivation of the scattering amplitude we
need to evaluate also the transition matrix element \cite{NS} of the
second term in (\ref{e2}), $\bra{0} g^2 \theta_{00}^{(G)}
\ket{\pi\pi}$. This tensor operator was discussed in the previous
Section, where we have found that it contributes a substantial
fraction, $8/23$, to the complete perturbative result. However, unlike
the scalar operator, the tensor term is not coupled to the anomaly,
and therefore $\bra{0} g^2 \theta_{00}^{(G)} \ket{\pi\pi} \sim g^2$,
where the coupling has to be evaluated at the heavy quarkonium
scale. This contribution therefore is of higher order in the strong
coupling, and will vanish in the heavy quark limit. Omitting it, we
come to the following low--energy expression,   
\begin{equation}
\bra{0}g^2 {{\bf E}^a}^2 \ket{\pi \pi}  = 
\left ( \frac{8 \pi^2}{b} \right ) q^2  + O(\alpha_s, m_\pi^2). \label{e2pi}
\end{equation}
Thus, in the heavy quark limit, the matrix element in question is
known up to $\alpha_s$ and $m_\pi^2$ corrections.

The result (\ref{e2pi}) has been derived in the chiral limit; the most
important correction coming from the finite mass of the pion is the
phase space threshold; we correct for it by writing down the spectral
density in the form ($q^2 \geq 4m_{\pi}^2$), 
\begin{eqnarray}
\rho^{\pi\pi}_{\theta}(q^2) &=&
\frac{3}{32\pi^2} \left ( \frac{q^2-4m_\pi^2}{q^2} \right )^{1/2} 
q^4,  \label{pimass}
\end{eqnarray}
which should be valid at small $q^2$. Substituting (\ref{pimass}) in
(\ref{yukawa}), we get the potential due to the $\pi\pi$ exchange; at
large $R$ 
\begin{equation}
V^{\pi \pi}(R) 
\rightarrow 
-\Big (\bar d_2 \frac{a_0^2}{\epsilon_0}\Big )^2
 \left ( \frac{4 \pi^2}{b} \right )^2
\frac{3}{2} (2 m_\pi)^4 
\frac{m_\pi^{1/2}}{(4\pi R) ^ {5/2} } 
e^{-2m_\pi R}.
\label{uncor}
\end{equation}
The same functional dependence of $\pi\pi$ exchange at large $R$ has
been obtained previously in Ref.~\cite{Peskin}, but up to an unknown
constant; in our approach, the overall strength of the interaction is
fixed. Note that, unlike the perturbative result (\ref{pertres}) which
is manifestly $\sim g^4$, the amplitude (\ref{uncor}) is $\sim g^0$ --
this ``anomalously" strong interaction is the consequence of scale
anomaly\footnote{Of course, in the heavy quark limit the amplitude
(\ref{uncor}) will nevertheless vanish, since $a_0 \to 0$ and
$\epsilon_0 \to \infty$.}.

The low--energy theorems\cite{nsvz,VZ} not only allow us to evaluate
explicitly the contribution of uncorrelated $\pi\pi$ exchange; they
also tell us that this contribution alone is not the full story
yet. Indeed, the $\pi\pi$ spectral density (\ref{pimass}) alone cannot
saturate the sum rule (\ref{let}) -- at high $\sigma^2$, the physical
spectral density approaches the spectral density of perturbation
theory, so the integral in (\ref{let}) does not get any contribution;
at small $\sigma^2$, the $\pi\pi$ spectral density (\ref{pimass}),
according to the chiral and scale symmetries is suppressed by $\sim
\sigma^4$. The low energy theorems {\it require} the presence of
resonant enhancement(s) \cite{MS} in the $0^{++}$ $\pi\pi$, and
perhaps multi-pion, $\bar{K}K$ and $\eta\eta$ channels as well. Here
we will leave this interesting problem aside, and study only the
influence of these resonances on the potential between the color
dipoles. To do this, we  introduce the pion scalar form factor
$F(q^2)$ and write down the spectral density as 
\begin{eqnarray}
\rho_{\theta}^{\pi\pi}(q^2) &=&
\frac{3}{32\pi^2} \left ( \frac{q^2-4m_\pi^2}{q^2} \right )^{1/2} 
q^4 |F(q^2)|^2. \label{densreal}
\end{eqnarray}
The form factor is directly related to the experimental $\pi\pi$ 
phase shifts by the Omn\`es--Muskhelishvili equation with the solution 
\begin{equation}
F(t) = 
\exp \left [\frac{t}{\pi}
\int ds \frac{\delta_0^0(s)}{s(s-t-i\epsilon)}\right ] \ ;
\label{omnes}
\end{equation}
with this formula we can make a full use of the experimental
information on the $\pi\pi$ correlation. For our calculation we have
used a simple analytic form~\cite{Ishida} for the phase shift
$\delta_0^0$ which was shown to fit the experimental data up to
$s_{\pi\pi} \simeq (1.2\ \rm{GeV})^2$. There are two main
contributions to the spectral density, $\rho^{\pi\pi}_\theta$, which
may be interpreted as the broad $\sigma$ and narrow $f_0$ resonances.

\begin{figure}[tb]
\centerline{\epsfxsize=0.7\textwidth \epsffile{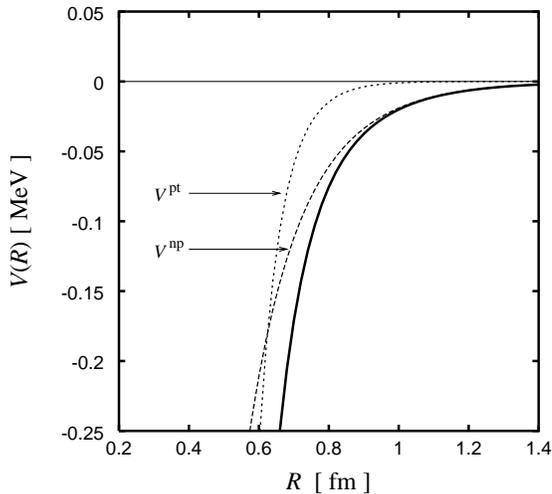}}
\caption{The potential between two $J/\psi$'s (bold solid line); 
the perturbative contribution $V^{pt}$ (dashed line) was evaluated within 
the invariant mass range $\sigma > 2\ \rm{GeV}$ in the spectral density;
$V^{np}$ (long--dashed line) is the non--perturbative contribution.  
}
\label{fig-pot1}
\end{figure}

In Fig.~\ref{fig-pot1} we show the resulting potential between two 
$J/\psi$ states. In computing it, we assumed that Coulomb relations
for the Bohr radius and the Rydberg energy $a_0=4/(3\ \alpha_s m)$ and
$\epsilon_0 = (4/3\ \alpha_s )^{-2} m=1/a_0^2\ m$ hold for the
$J/\psi$. Using as an input $\epsilon_0 = 2 M_D-M(J/\psi)$=642 MeV and
$m$=1.5 GeV, we get $\alpha_s$=0.87 and $a_0$=0.20 fm.

It can be clearly seen from Fig.~\ref{fig-pot1} that at large distances 
the non--perturbative interaction dominates over the perturbative one. 
To evaluate the amplitude, we had to use an experimental input -- 
the $\pi\pi$ phase shifts, and the detailed shape of the potential
depends on this input. However, as we will now see, the dominance of
the non--perturbative interaction is a model--independent consequence
of the low--energy theorems. Moreover, its overall strength can be
shown to be completely determined by the energy density of
non--perturbative vacuum of QCD.

\section{The Sum Rule}

Consider the integral over the non--perturbative part of the potential; 
according to (\ref{yukawa}), it can be written down as
\begin{eqnarray}
&&
\int_a^\infty d^3{\bf R} \left( V(R) - V^{pt}(R) \right)
\nonumber \\
& =& 
  -
\Big ( \bar d_2 \frac {a_0^2}{\epsilon_0} \Big )^2 
\Big ( \frac{4 \pi^2 }{b}\Big ) ^2 
\int d \sigma^2  \left( \rho (\sigma^2) - \rho^{pt}(\sigma^2) \right) 
\int_a^\infty d R R^2  
\frac{1}{R}e^{-\sigma R}
\nonumber \\
&=& -
\Big ( \bar d_2 \frac {a_0^2}{\epsilon_0} \Big )^2 
\Big ( \frac{4 \pi^2 }{b}\Big ) ^2
\int \frac{d\sigma^2}{\sigma^2}
\left( \rho (\sigma^2) - \rho^{pt}(\sigma^2) \right) 
\Gamma(2,\sigma a).
\label{sr1}
\end{eqnarray}
As a consequence of asymptotic freedom, $\rho (\sigma^2) 
\to \rho^{pt}(\sigma^2)$ at high $\sigma$, so the integral in (\ref{sr1}) 
attains its value in the region of $\sigma < \sigma_0$, where $\sigma_0$ 
is some characteristic scale at which the perturbative regime sets in. 
In the heavy quark limit the size of quarkonium 
$a \sim (\alpha_s m)^{-1} \to 0$, and when 
$\sigma_0 a \ll 1$, $\Gamma(2,\sigma a) \simeq 1$ in the entire region 
of integration. The integral in (\ref{sr1}) then, up to perturbative 
corrections $\sim O(g^4)$, coincides with the integral in (\ref{let}). 
Therefore we can re--write (\ref{sr1}), in the heavy quark limit, 
as a sum rule
\begin{equation}    
\int d^3{\bf R} \left( V(R) - V^{pt}(R) \right) = 
-\Big ( \bar d_2 \frac {a_0^2}{\epsilon_0} \Big )^2 
\Big ( \frac{4 \pi^2 }{b}\Big ) ^2 \ 16 \left|\epsilon_{vac}\right|, 
\label{sr2}
\end{equation}
which expresses the overall strength of the interaction between 
two color dipoles in terms of the energy density of the 
non-perturbative QCD vacuum.

\section{Final Remarks}

What are the implications of our results? 
First, the pion clouds which dominate interactions of small color
dipoles at low energies, as revealed by our analysis, may as well be
important in high--energy scattering; this was suggested long time ago
on general grounds \cite{AG,Gribov3}. However it is not yet clear if
one can extend our approach to the scattering at high
energies\footnote{For a very interesting related discussion see
Ref.~\cite{Bjorken}.}.

Second, the fact that pions (and therefore light quarks) dominate the
long--distance interactions of heavy quark systems is important for
the lattice QCD simulations. Our findings suggest that to extract the
information on the properties and interactions of heavy quarkonia from
lattice QCD one should go beyond the ``quenched" approximation.

\section*{Acknowledgments}

We thank M. Gyulassy, T.D. Lee, L.D. McLerran, A.H. Mueller, 
R.D. Pisarski, E.V. Shuryak and M.B. Voloshin for useful discussions.
D. K. is also grateful to {\mbox{ J. Ellis}} for explanations of  
scale invariance and to H. Mishra and H. Satz for  
collaboration on related topics. He thanks the Organizers 
for their invitation to this memorable Workshop.

\end{document}